\documentclass[10pt,a4paper,english,prb, twocolumn]{revtex4-2}

\usepackage[T1]{fontenc}
\usepackage[utf8]{inputenc}
\usepackage{amsmath}
\usepackage{graphicx}

\makeatletter

\pdfpageheight\paperheight
\pdfpagewidth\paperwidth

\makeatother

\usepackage{babel}
\begin{document}
\title{Infinite magneto-resistance and bipolar effect in spin valves driven
by spin batteries}
\author{Khuôn-Viêt Pham}
\affiliation{Université Paris-Saclay, CNRS, Laboratoire de Physique des Solides,
91405 Orsay, France}
\begin{abstract}
It is shown that spin valves under suitable symmetry conditions exhibit
an ON-OFF response to a spin battery. While a spin valve driven by
a charge battery displays the usual GMR (Giant Magneto-Resistance),
a pure spin current or pure spin accumulation can generate an infinite
magneto-resistance effect(IMR). In practice the effect is curtailed
by asymmetry but the achievable magneto-resistance ratio is still
predicted to be unusually large in several example setups. It is closely
related to the bipolar effect (BE) of Johnson transistor which can
be triggered under the same symmetry conditions.
\end{abstract}
\maketitle

\section{Introduction}

This paper introduces a new magneto-resistive effect, the infinite
magneto-resistance (IMR) related to the giant magneto-resistance spin
valve (GMR \citep{dieny_giant_1991,chappert_emergence_2007}) and
to the bipolar spin transistor (BST) of Johnson \citep{johnson_bipolar_1993,johnson_spin_1993}. 

The IMR does not rely on specific materials such as half-metals to
increase GMR (Heusler alloys as in Kimura et al\citep{kimura_room-temperature_2012})
or coherent tunneling through MgO barriers \citep{mathon_theory_2001,butler_spin-dependent_2001}.
It consists in achieving a zero signal for a system with 2 magnetically
active parts (like two ferromagnetic layers in a spin valve) so that
the optimistic (resp. pessimistic) MR ratio tends to infinity (resp.
unity), hence the naming of the effect as IMR. Its main characteristics
are: (1) the use of a spin battery instead of a charge battery; (2)
the reliance on symmetry to get rid of the offsets preventing a bipolar
response in Johnson BST; (3) the reliance on Johnson-Silsbee charge-spin
coupling \citep{johnson_coupling_1988} to measure the effect. Of
course, the effect is curtailed by asymmetries which always exist
so that the MR ratio can never be infinite and it will be one of the
objectives of the paper to quantify to impact of slight asymmetries
and show that the MR ratio can still be very large.

The starting point of our proposed effect is the BST. In the latter,
one has a spin valve like structure $F1-N-F2$ with two collinear
ferromagnets $F1$ and $F2$ connected by a paramagnetic metal $N$.
A charge current flows through the first ferromagnet $F1$ but NOT
through the second $F2$. Thanks to spin injection it induces a spin
current and spin accumulation which spill into the second ferromagnet
$F2$. The spin accumulation in $F2$ is detected as a charge voltage
through Johnson-Silsbee charge spin coupling. It was predicted that
the signal would be bipolar as a function of the magnetization polarization
(up or down): $V_{c}\propto s\mathrm{gn}\left(M_{z}\right)\propto\pm V_{1}$.
Experimentally the bipolarity has proved elusive to detect and one
observes instead 
\begin{equation}
V_{c}=V_{0}+V_{1}\,\sigma
\end{equation}
 (where $\sigma=\pm$) with a \textbf{voltage offset} $V_{0}$ \citep{johnson_calculation_2007,fert_semiconductors_2007,johnson_bipolar_1993,ichimura_geometrical_2004,hamrle_current_2005,hamrle_three-dimensional_2005,garzon_temperature-dependent_2005,casanova_control_2009,bakker_interplay_2010}.
The origin of these offsets have been credited to various causes such
as current inhomogeneities \citep{ichimura_geometrical_2004}, heating
\citep{garzon_temperature-dependent_2005,casanova_control_2009,bakker_interplay_2010},
departures from one dimensional geometry \citep{johnson_calculation_2007}.

Conceptually one can deconstruct the BST into three components: (1)
a spin battery, made up of the ferromagnet $F1$ when a charge current
flows through it; (2) a load $F2$ for the spin current and spin accumulation
created by the spin battery; (3) a charge (voltage or current) measurement
apparatus.

The systems we consider will be driven by a spin battery like the
original BST but differ in that the load is not a single ferromagnet
but has at least two magnetically active regions. For instance for
collinear systems, we will have in general a voltage response 
\begin{equation}
V_{c}=V_{0}+V_{1}\,\sigma_{1}+V_{2}\,\sigma_{2}+V_{12}\,\sigma_{1}\,\sigma_{2}
\end{equation}
 where $\sigma_{1/2}=\pm$ track the magnetization directions. The
voltage offsets come from $V_{0}$ and $V_{12}$. But should they
vanish due to some some symmetry, one would then have 
\begin{equation}
V_{c}=V_{1}\,\sigma_{1}+V_{2}\,\sigma_{2}
\end{equation}
 exhibiting a bipolar response when both magnetizations are switched
($\sigma_{1/2}\longrightarrow-\sigma_{1/2}$). Suppose now that due
to a stronger symmetry $V_{1}=V_{2}$ (resp. $V_{1}=-V_{2}$); $V_{c}$
will then obviously vanish whenever the two magnetizations are antiparallel
(resp. parallel). This implies that the optimistic MR ratio $\left(V_{HIGH}-V_{LOW}\right)/V_{LOW}$
will be infinite. This is the IMR which the main topic of this paper.
In such a spin valve, a single magnetization needs to be switched;
if instead both are switched, one recovers a BE.

Aside from symmetry, the other main ingredient is the use of a spin
battery instead of a regular charge battery. The reason is that oftentimes
the response $r_{c}\left(\sigma\right)=\partial V_{c}/I_{c}$ is even
in $\sigma$ (the magnetization direction), $r_{c}\left(-\sigma\right)=r_{c}\left(\sigma\right)$
which implies that $V_{1}=0$ so that the IMR is not possible with
a pure charge battery (since it lacks any magnetically active region).
In contrast a spin battery will induce an odd term in the voltage
response, which is required for the BE and IMR.

The third ingredient is the reliance on Johnson-Silsbee charge-spin
coupling (the coupling of charge and spin transport in magnetic systems)
to get a (charge) voltage or current reading from magnetic changes.
The advantage of a voltage measurement over a current one is that
it avoids Joule heating. But theoretically either one could be done.

\begin{figure}[h]
\noindent \centering{}\includegraphics[width=1\columnwidth]{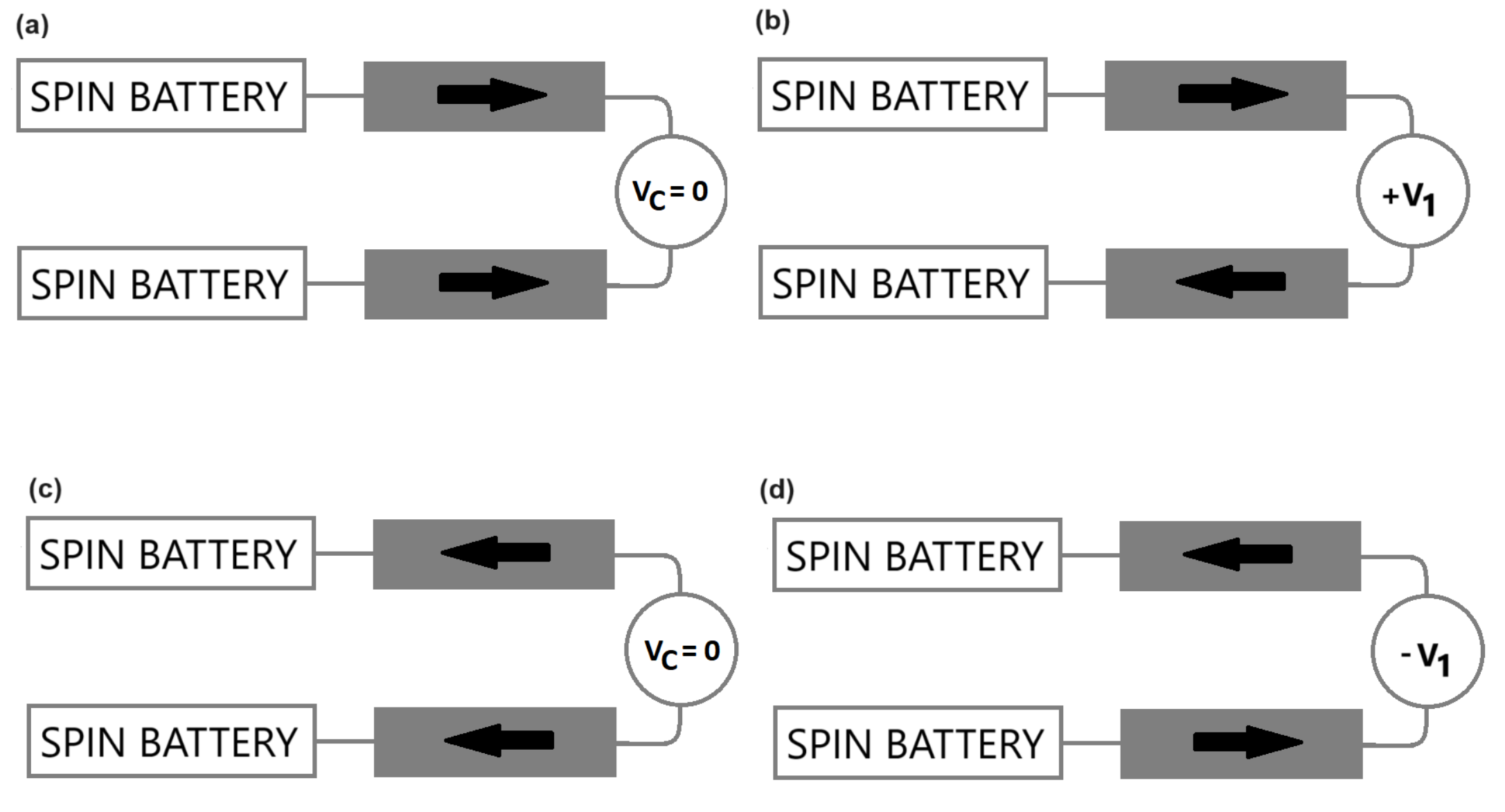}\caption{(a) The voltage measured across two copies of the same system (and
batteries) vanishes by symmetry. (b) If one ferromagnet is flipped,
one gets a non-vanishing voltage. (c) If both ferromagnets are flipped
from (a), the voltage still vanishes. (d) The voltage is now the opposite
of (b) since (d) follows from (b) by flipping both magnetizations,
which amount to exchanging both systems. \label{fig:twin-initial}}
\end{figure}

\begin{figure}[h]
\noindent \centering{}\includegraphics[width=1\columnwidth]{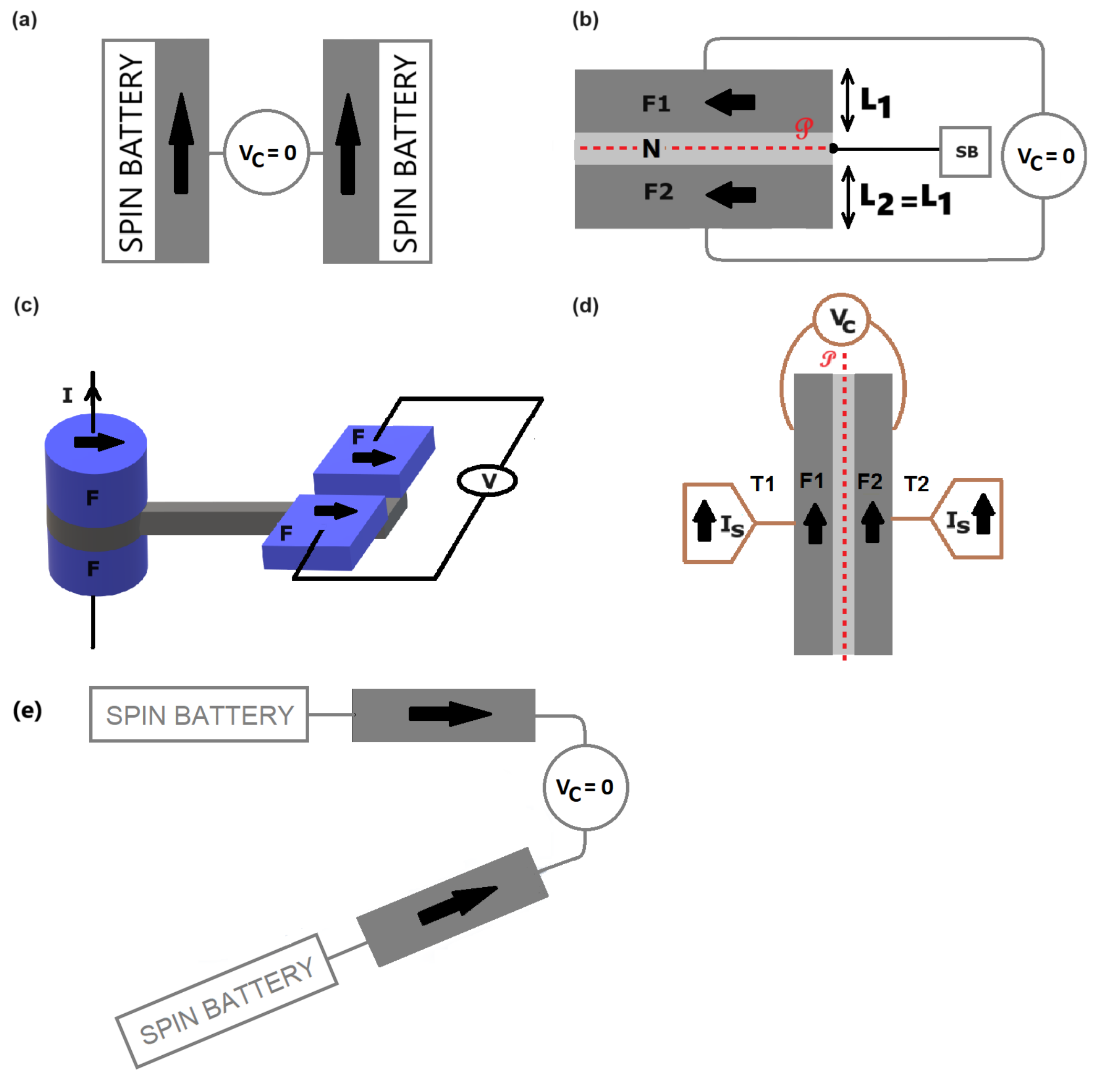}\caption{(Color online) (a) The voltage measured across two copies of the same
system (and batteries) vanishes by symmetry. (b) FNF valve connected
to a spin battery and a voltmeter. If the combined system is symmetrical
under left-right exchange, then the voltage must vanish by symmetry.
The layers can be viewed as parallel to the figure plane (as in non-local
geometries) or to be perpendicular. SB: spin battery. (c) Same as
(b) but depicted with an FNF nanopillar playing the role of the SB.
(d) CPP trilayer driven symmetrically by two spin batteries. For parallel
magnetizations, the voltage again vanishes (for symmetric probes).
(e) Twin system with an angular offset. The black arrows depict the
magnetization directions. \label{fig:twin}}
\end{figure}

What is of paramount importance is to have symmetrical setups which
will allow to get rid by construction of any offsets. A simple recipe
is to have two copies of the same system as in Fig. \ref{fig:twin-initial}-a:
the offsets obviously cancel out. This is an example of what we will
christen as 'twin systems'. In Fig. \ref{fig:twin-initial}, the ferromagnets
can be thought of being in a non-local geometry. But they could also
be thin films as in Fig. \ref{fig:twin} -a

Another example of a 'twin system' is given in Fig. \ref{fig:twin}-b.
It is an $FNF$ valve in a non-local geometry with identical ferromagnets
and layers parallel to the figure plane; it could also be considered
to be a trilayer with layers perpendicular to the figure plane. A
spin battery injects a spin current through the intermediate $N$
layer. The charge voltage across the trilayer obviously vanishes when
the ferromagnets have parallel magnetizations but is non-zero for
anti-parallel ones.

In practice the vanishing of the signal is constrained by the degree
of symmetry. Suppose for instance that the widths of the ferromagnets
are equal to $l_{m}\pm\delta l$. Then we expect a scaling: 
\begin{equation}
\frac{V_{c}^{AP}-V_{c}^{P}}{V_{c}^{P}}\sim\frac{l_{m}}{\delta l}
\end{equation}
so for widths equal with an accuracy of $1\%$, the ratio reaches
$10^{2}=10^{4}\:\%$. Other geometrical asymmetries will of course
affect this rough estimate\emph{.} For the same twin setup, the BE
is achieved by starting from an antiparallel configuration and flipping
both magnetizations.

There is a wide variety of spin batteries currently in use and among
others: spin injection devices (a charge current gets polarized by
traversing a ferromagnet \citep{julliere_tunneling_1975,baibich_giant_1988,binasch_enhanced_1989});
spin Hall devices (a transverse spin current is generated when there
is a longitudinal charge current due to spin-orbit interactions \citep{valenzuela_direct_2006});
spin pumps (a rotating bulk magnetization engenders a spin current,
as in the ferromagnetic resonance battery \citep{brataas_spin_2002,watts_unified_2006});
tunnel junctions relying on photoexcitation (circularly polarized
light at FM/semiconductor tunnel junctions excites spin polarized
carriers \citep{prins_spin-dependent_1995}); devices relying on the
spin Seebeck effect (a thermal gradient can generate a spin current)
\citep{uchida_observation_2008,kirihara_flexible_2016}. Fig. \ref{fig:twin}-c
shows the twin system from Fig. \ref{fig:twin}-b with an FNF nanopillar
acting as a spin battery.

There are other possible setups than 'twin setups' which will exhibit
IMR and bipolarity, as a result of symmetry. For instance consider
a CPP symmetric trilayer where instead of a charge battery, one connects
the layers to two spin batteries as in Fig. \ref{fig:twin}-d. (By
CPP, current perpendicular to plane, we mean that if one replaced
the voltage probes by current probes, the charge current would flow
perpendicular to the layers.) This setup exhibits IMR for identical
incoming spin currents and parallel magnetizations (parallel IMR).
We will discuss later a similar setup with opposite spin currents;
one then has IMR for antiparallel magnetizations (AP IMR). Note also
that, as shown on \ref{fig:twin}-e, the two copies of the same system
can have arbitrary relative orientations provided the voltage measurement
probes remain symmetric.
\begin{figure}[h]
\noindent \centering{}\includegraphics[width=0.8\columnwidth]{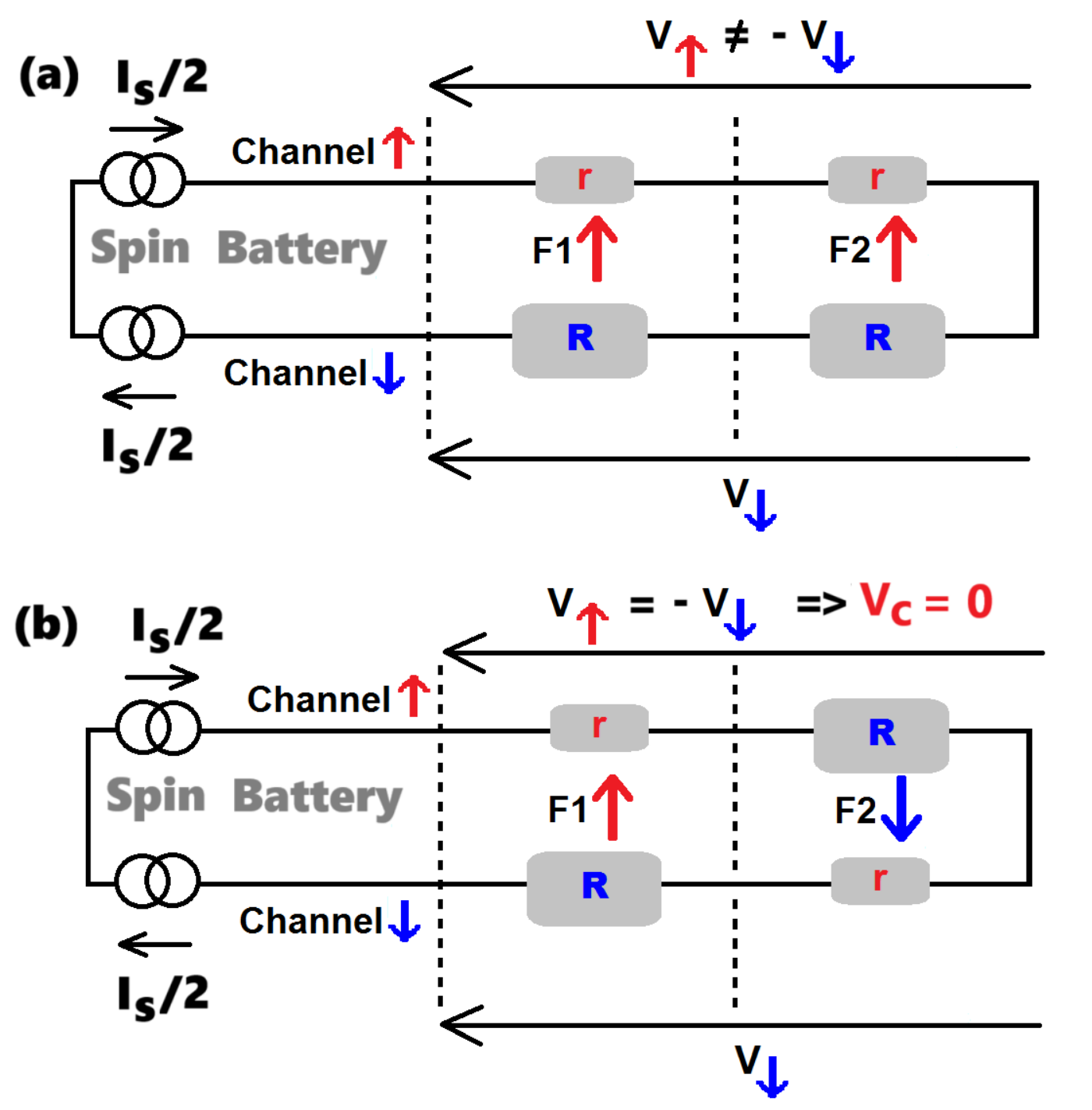}\caption{(Color online) Resistor model of a spin valve driven by a spin current
source. The charge voltage is $V_{c}=\left(V_{\uparrow}+V_{\downarrow}\right)/2$.
(a) $V_{c}\protect\neq0$ when ferromagnets $F1$ and $F2$ have parallel
magnetizations. (b) But for anti-parallel magnetizations, the voltage
$V_{c}$ obviously vanishes since $V_{\uparrow}=\left(r+R\right)\:I_{s}/2=-V_{\downarrow}$.
\label{fig:imr-1}}
\end{figure}

\begin{figure}
\noindent \centering{}\includegraphics[width=1\columnwidth]{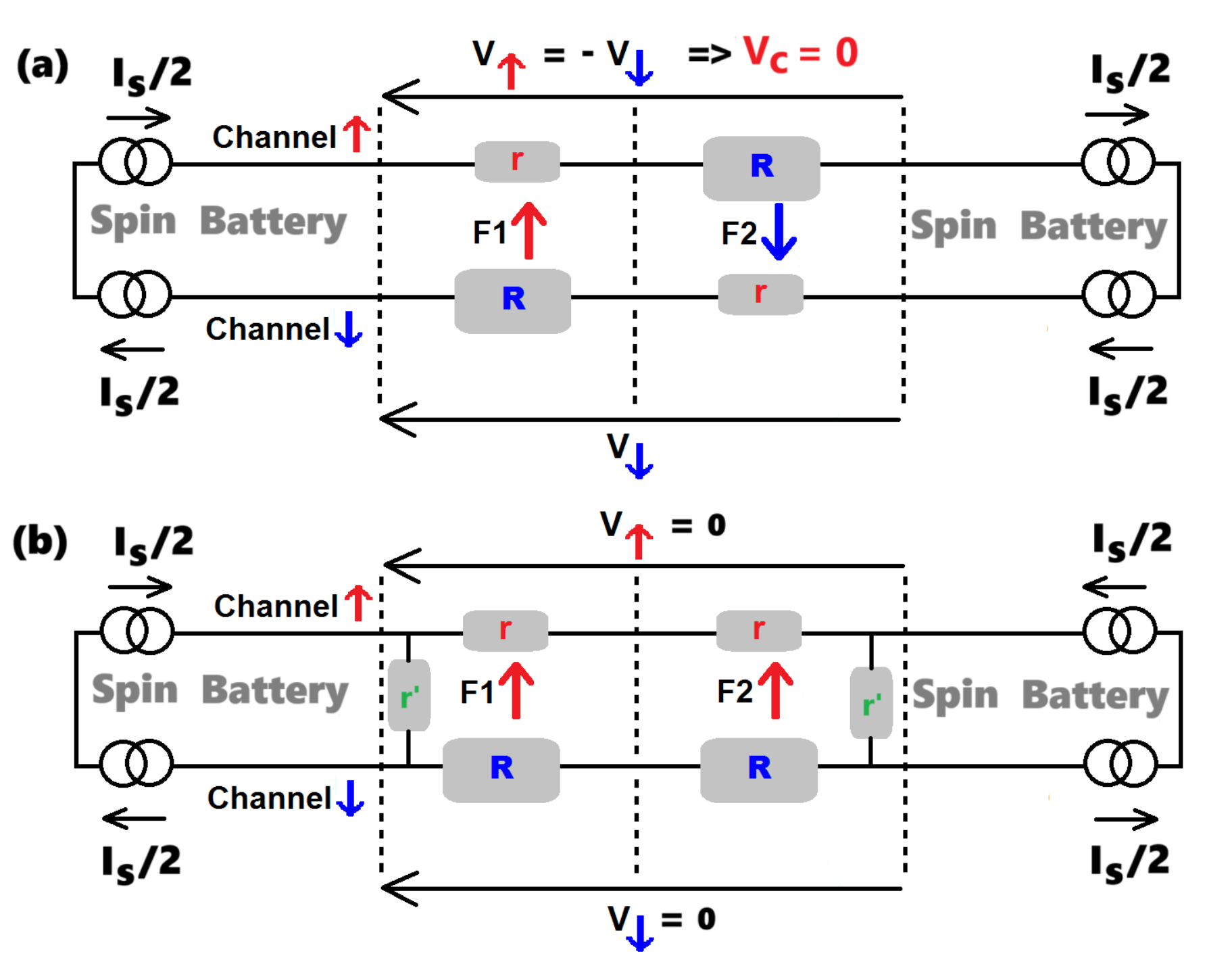}\caption{(Color online) (a) Resistor analogy for AP IMR of CPP spin valve connected
to two spin batteries. Since spin flip is neglected, the spin current
source on the right can be removed because it is redundant. (b) The
resistor analogy for parallel IMR of CPP spin valve connected to 2
spin current sources requires addition of spin mixing either in bulk
and/or at the interface between the ferromagnets. In the resistor
circuit, the resistance $r'$ ensures the current to be in a closed
loop while no current flows through resistances $r$ and $R$. Therefore
the charge voltage $V_{c}=\left(V_{\uparrow}+V_{\downarrow}\right)/2$
vanishes. The same conclusion is reached by adding a mixing resistance
at the interface between the ferromagnets.\label{fig:P-IMR}}
\end{figure}

The AP IMR for CPP spin valve is actually easy to understand in terms
of a resistor analogy with a single spin battery (due to current conservation
on each spin channel, a second spin battery is irrelevant, see Fig.\ref{fig:P-IMR}-a).
We consider a spin current source in a two-channel model ( Fig. \ref{fig:imr-1})
: this is tantamount to having opposite charge current sources for
each spin channel. In the resistor analogy for a CPP symmetric spin
valve with anti-parallel magnetizations \emph{driven by a charge battery},
the currents flowing in each spin channel would be equal $I_{\uparrow}=I_{\downarrow}$
and the voltages for each spin channel would be identical in the anti-parallel
case $V_{\uparrow}=V_{\downarrow}$ . In the antiparallel case for
a spin current source driving the resistors ( Fig.\ref{fig:imr-1}-(b)
), since the currents are opposite $I_{\uparrow}=-I_{\downarrow}=I_{s}/2$,
the channel voltages are also opposite $V_{\uparrow}=-V_{\downarrow}$
so that $V_{c}=0$. The charge voltage divided by the spin current
has the dimension of a resistance $R=V_{c}/I_{s}$. The corresponding
MR ratio $\left(R^{AP}-R^{P}\right)/R^{AP}$ goes to infinity (IMR
effect). A similar resistor analogy can be drawn for parallel IMR
for CPP spin valves but requires to take into account spin flips and
transfer from one spin channel to the other. See Fig.\ref{fig:P-IMR}-b.

\section{Response function, BE \& IMR.}

\subsection{Trans-resistance.}

Let us define a charge-spin trans-resistance encoding the response
to a spin battery:
\begin{equation}
r_{cs}=\left(\frac{\partial V_{c}}{\partial I_{s}}\right)_{I_{c}=0},
\end{equation}
where the spin current is defined in electrical units as $I_{s}=I_{\uparrow}-I_{\downarrow}$
for collinear setups. It has the dimension of a resistance. The basic
geometries we consider are therefore 3 terminal ones, two terminals
used for the voltage measurement and one terminal connected to the
spin battery. We will make the hypothesis that the spin battery is
a perfect spin current source which outputs a constant spin current
$I_{s}$, whatever the load it is attached to. This is not a very
stringent condition. It is easy to take into account spin backflow
into the battery through the internal spin resistance concept \citep{pham_internal_2018},
which in practice will translate into a reduced response. Let us suppose
that the load comprises $N=2$ magnetically active regions, whose
\textbf{magnetizations are collinear (but maybe have different directions)}
and can be controlled independently. One can write therefore: $r_{cs}=r_{cs}\left(\sigma_{1},\;\sigma_{2}\right)$
where $\left\{ \sigma_{1}=\pm1,\;\sigma_{2}=\pm1\right\} $ index
the magnetizations of the $2$ regions relative to a given quantization
axis. $r_{cs}$ is simply a multilinear polynomial in $\sigma_{1}$
and $\sigma_{2}$: 
\begin{equation}
r_{cs}\left(\sigma_{1},\,\sigma_{2}\right)=r_{0}+r_{1}\,\sigma_{1}+r_{2}\,\sigma_{2}+r_{12}\,\sigma_{1}\,\sigma_{2}.
\end{equation}
 The resistance $r_{cs}$ can take four independant values.

\subsection{Bipolar effect (BE). }

This requires the concurrent switching of both magnetizations. One
has a bipolar effect if for some values of $\sigma_{1}$ and $\sigma_{2}$
\begin{equation}
r_{cs}\left(-\sigma_{1},-\sigma_{2}\right)=-r_{cs}\left(\sigma_{1},\sigma_{2}\right);\label{eq:reversal-1}
\end{equation}
which implies that $r_{0}=-r_{12}\,\sigma_{1}\,\sigma_{2}.$ In the
special case $r_{0}=0,\;r_{12}=0,$ the bipolarity will occur for
any values of $\sigma_{1}$ and $\sigma_{2}$. The twin system of
Fig.\ref{fig:twin}-a obeys 

\begin{equation}
r_{cs}\left(\sigma_{2},\sigma_{1}\right)=-r_{cs}\left(\sigma_{1},\sigma_{2}\right)\label{eq:cond}
\end{equation}
so that $r_{0}=0,\;r_{12}=0,$$r_{1}=-r_{2}$. For Fig. \ref{fig:twin}-b,
eq.(\ref{eq:cond}) is also true since a symmetry through plane $(\mathcal{P})$
parallel to the layers and perpendicular to the Figure maps each ferromagnetic
layer unto the other but leaves the spin battery invariant. But the
voltage probes are exchanged across that plane so that $V_{c}\longrightarrow-V_{c}$.
It follows that: $r_{cs}\left(\sigma_{1},\,\sigma_{2}\right)=r_{1}\cdot\left(\sigma_{1}-\sigma_{2}\right).$
If one starts from an antiparallel configuration and swaps both magnetizations,
one therefore has a bipolar response: 
\begin{equation}
r_{cs}\left(\sigma_{1},\,-\sigma_{1}\right)=2r_{1}\sigma_{1}=-r_{cs}\left(-\sigma_{1},\,\sigma_{1}\right).
\end{equation}

\subsection{IMR. }

For the twin system of Fig.\ref{fig:twin}-a \& b, it also follows
from eq.(\ref{eq:cond}) that $r_{cs}^{P}=r_{cs}\left(\sigma_{1},\,\sigma_{1}\right)=0$
while $r_{cs}^{AP}\neq0$. We have a \textbf{parallel IMR}. The pessimistic
MR ratio is 100\% and the optimistic MR ratio is infinite. If one
starts from an antiparallel setup and one switches a single magnetization,
the response vanishes (IMR). A similar symmetry condition can not
be enforced when there is a single $N=1$ magnetized region because
in order to detect a charge voltage through Johnson-Silsbee charge-spin
coupling, the measurement probes must be positioned asymmetrically.

Since the condition for IMR also allows the BE, let us compare the
two effects for the setups of Fig. \ref{fig:twin}-a, \ref{fig:twin}-b.
IMR is observed by switching a single magnetization while bipolarity
requires switching two magnetizations, which may or may not be possible
or desireable in a given setup. The amplitude of the IMR effect is
$\left|r_{cs}^{AP}-r_{cs}^{P}\right|=2r_{1}$, while the bipolar effect
has an amplitude which is twice as large. There is therefore a trade-off
between the two effects: IMR is smaller but easier to observe, requiring
a single magnetization to be switched, while bipolarity has the larger
signal but requires both magnetizations to be switched.

Let us compare with the regular charge resistance $r_{c}=\left(\frac{\partial V_{c}}{\partial I_{c}}\right)$;
the latter has a similar dependence in general. But if the system
is invariant when both magnetizations are reversed, one has the simpler
dependence: $r_{c}\left(\sigma_{1},\,\sigma_{2}\right)=r_{c0}+r_{c12}\,\sigma_{1}\,\sigma_{2}.$
One recovers the usual GMR for a spin valve subjected to a charge
battery. An infinite MR ratio is much more difficult to engineer.
For instance, let us consider Juilliere model for TMR. Then: $r_{c}\propto\left(1+P_{1}P_{2}\right)$
where $P_{1/2}$ is a DOS spin polarization. For identical layers,
$\left|P_{1}\right|=\left|P_{2}\right|=P$. An infinite MR ratio would
then require $\left(1-P^{2}\right)=0$, that is $P=1$ or a half-metal.
This is a very stringent condition when compared with the IMR discussed
above.

In GMR spin valves, one usually anchors one of the two ferromagnets
so that a single ferromagnet can be switched under normal operating
conditions. Doing so here would obviously prevent the symmetries enabling
the IMR and bipolar effects. This is another difference with conventional
GMR spin valves. The switching field needs to be tailored according
to the effect one wishes to trigger, whether it be IMR or the bipolar
effect.

\section{Example setups.}

\subsection{Twin setup with shared spin battery.}

To quantify the impact of asymmetries, one can make explicit calculations
for the geometry of Fig.(\ref{fig:twin}-b) in the one dimensional
limit, using Valet-Fert drift diffusion model. We introduce $P_{F1}$
and $P_{F2}$ are the conductivity polarizations for ferromagnets
$1$ \& $2$, contact resistance $r_{c}$, conductance polarization
$P_{c1}$ and $P_{c2}$ at the interfaces between the ferromagnets
$1$ \& 2 and the central paramagnetic layer, contact resistances
$r_{c}'$, conductance polarization $P_{c1}'$ and $P_{c2}'$ at the
interfaces between the ferromagnets $1$ \& 2 and the measurement
leads. We also define $l_{1/2}$ the ferromagnet layer widths relative
to $l_{F}$ the spin relaxation length of the ferromagnets. Defining
$l_{m}=\frac{l_{1}+l_{2}}{2},\;\delta l=\frac{l_{1}-l_{2}}{2}$ as
well as $\sinh_{m}=\sinh l_{m},\;\cosh_{m}=\cosh l_{m},$
\begin{align}
X & =\frac{r_{N}+r'_{c}}{r_{F}},\:f_{m}=\frac{\sinh_{m}+X\:\cosh_{m}}{\cosh_{m}+X\:\sinh_{m}},
\end{align}
one finds after tedious calculations that $r_{cs}$ splits into 3
parts, a bulk contribution (I) and two interfacial ones (II \& III):
\begin{align}
r_{cs}^{I} & =\left(P_{F1}-P_{F2}\right)\:\left[-f_{m}\left(\cosh_{m}-1\right)+\sinh_{m}\right]\:\frac{r_{F}}{2}\nonumber \\
 & +\delta l\:\left(P_{F1}+P_{F2}\right)\:\frac{r_{F}}{2}\:\alpha,\\
r_{cs}^{II} & =\left(P_{c1}-P_{c2}\right)\:\frac{r_{c}}{2}+\delta l\:\left(P_{c1}+P_{c2}\right)\:\frac{r_{c}}{2}\:\beta,
\end{align}
where $\alpha$ and $\beta$ are factors which depend on the geometry
and the various spin resistances, namely
\begin{align}
\alpha & =\left[\cosh_{m}-\left(f_{m}^{2}-1\right)\left(\cosh_{m}-1\right)-f_{m}\sinh_{m}\right]\nonumber \\
 & +\frac{f_{m}^{2}-1}{\frac{r_{c}}{r_{F}}+f_{m}}\left[-f_{m}\left(\cosh_{m}-1\right)+\sinh_{m}\right],\\
\beta & =\frac{f_{m}^{2}-1}{\frac{r_{c}}{r_{F}}+f_{m}}.
\end{align}
\begin{align}
r_{cs}^{III} & =\left(P_{c1}'-P_{c2}'\right)\:\frac{r_{c}'}{2}\:\left(\cosh_{m}-f_{m}\sinh_{m}\right)\\
 & +\delta l\:\left(P_{c1}'+P_{c2}'\right)\:\frac{r_{c}'}{2}\:\left(\cosh_{m}-f_{m}\sinh_{m}\right)\nonumber 
\end{align}
One can compute a resistance ratio as $MR=\frac{R_{AP}-R_{P}}{R_{P}}$
where the subscripts $P/AP$ refer to parallel or anti-parallel magnetizations;
it follows that:
\begin{equation}
R_{P}=\circ\left(\delta l\right),\qquad MR\propto\delta l^{-1}
\end{equation}
Asymmetries in transverse directions should in principle be taken
into account so that most generally one expects
\begin{equation}
MR\propto\left(\sum_{i=x,y,z}C_{i}\:\frac{\delta l_{i}}{l_{i}}\right)^{-1}
\end{equation}
but for thin films, the dominant contribution will come from the width
so that other asymmetries can be neglected. As an illustration, consider
a Co-Cu-Co trilayer. For all the following figures, we refer to \citep{bass_current-perpendicular_1999};
Piraux et al find at $77\:K$ that $l_{F}=36\:nm$. MSU Reilly et
al have data consistent with a large $l_{F}$ maybe larger than $60\:nm$
at $4.2\:K$. If the widths are about $10\pm0.5\,nm$, $\delta l\sim\frac{1}{60}$
so up to numerical factors $MR\sim6000\%.$

It is also important to have an idea of the range of the HIGH signal
(the LOW signal being close to 0). Let's use MSU results for CoCu
at $4.2\:K$ and for simplicity assume that $P_{c}\sim P_{c}'=0.77\pm0.05$,
$P_{F}=0.46\pm0.05,$ $2r_{c}A\sim2r_{c}'A=1.02\pm0.04\:f\Omega m^{2}$,
$t_{Co}=2-10\,nm,\:l_{Co}>60\:nm$ and $\rho_{F}(Co)=75\pm5\:n\Omega m,$
$\rho_{N}(Cu)=6\pm1\:n\Omega m.$ Then $X\sim1$, $f_{m}\sim1$. So
for AP ferromagnets: 
\begin{equation}
r_{cs}^{high}\,A\sim1\:f\Omega m^{2}.
\end{equation}
This will be in the range of $R=1\:\Omega$ for $A=10^{-3}\mu m^{2}$
which corresponds to a nanopillar.

\subsection{Twins with individual spin batteries. }

We consider two identical systems connected to a spin battery and
interconnected by voltage probes (Fig.(\ref{fig:twin}-a)). This geometry
is better suited to the observation of IMR than the previous one because
it is easier to switch a single ferromagnet since they can be well
separated spatially. One must take into account spin batteries asymmetry
$\delta I_{s}$; one finds a MR ratio 
\begin{equation}
R_{P}=\circ\left(\delta l\right)+\circ\left(\delta I_{s}\right),\qquad MR\propto\left(\frac{\delta l}{l}+C\,\frac{\delta I_{s}}{I_{s}}\right)^{-1}.
\end{equation}
For very wide ferromagnets, it can be shown that it is better to have
$X=\frac{r_{N}+r'_{c}}{r_{F}}\sim1$ so that $C\sim1-2$. In the opposite
limite of very thin ferromagnets, $C\propto X$, the MR ratio will
be boosted by transparent contacts ($r_{c}'\ll\left(r_{N},\;r_{F}\right)$)
which lower $C$. For very resistive contacts such that $\left(r_{c}',\:r_{c}\right)\gg\left(r_{N},\;r_{F}\right)$
and $X\gg1$, one finds that $C\propto1/X$ provided the ferromagnets
are not too thin. The reason for this potential enhancement is that
the resistive contacts hinder the backflow of spin accumulation and
allow its build-up. But the spin current asymmetry will likely limit
any such improvement. The HIGH signal is dominated by interfaces in
the limit of very thin ferromagnets: $r_{cs}^{high}\longrightarrow2\:\left\{ P_{c}\,r_{c}+P_{c}'\,r_{c}'\right\} $
while for very thick layers: $r_{cs}^{high}\longrightarrow2\:\left\{ P_{c}\,r_{c}+P_{F}\,r_{F}\right\} .$
Using the same parameters as above, one finds again a range $r_{cs}^{high}\,A\sim1\:f\Omega m^{2}.$

\subsection{Symmetric CPP spin valves operated by two spin batteries. Parallel
IMR. }

One can define new transport coefficients per:
\begin{equation}
V_{c}=r_{cs}^{L}\:I_{s,L}+r_{cs}^{R}\:I_{s,R}.
\end{equation}

Since there is a new control parameter, it is easy to get a zero response
(IMR) by adjusting one spin current relative to the other. One can
then sweep the value of, say, the spin current $I_{s,L}$ until $I_{s,L}=-\frac{r_{cs}^{R}\:I_{s,R}}{r_{cs}^{L}}.$\textbf{
No symmetry is required} but the trade-off is that (1) the system
is larger since there are two spin batteries, (2) while an adjustment
is easy to do in a lab context, it is much less convenient to do so
in industrial applications.

It is therefore advantageous to engineer symmetrical systems which
can have reproducible behaviours. Consider for instance Fig.\ref{fig:twin}-d
which shows a spin valve in a CPP geometry (if the voltmeter was shorted,
a charge current would flow in a direction orthogonal to the layers).
\textbf{Note that the setup in Fig. \ref{fig:twin}-d could represent
also a nano-pillar or a non-local geometry in addition to the usual
multi-layer thin films.} The spin current from the batteries are oriented
to be \textbf{incoming in the load} (outgoing from the batteries).
If the ferromagnetic layers and the spin batteries are identical,
there is a plane of symmetry $\mathcal{P}$ in the paramagnetic layer
parallel to the layers. Obviously $V_{c}=0$ when the magnetizations
are parallel. One has a \textbf{parallel IMR}. If one starts with
antiparallel magnetizations and switches both of them, one gets a
bipolar effect. See Fig.\ref{fig:P-IMR} for the resistor analogy.

After rotating the setup around an axis in the middle of the paramagnetic
layer:
\begin{align}
V_{c}\longrightarrow-V_{c}, & \:I_{s,L}\longrightarrow I_{s,R},\:\left(\sigma_{1},\:\sigma_{2}\right)\longrightarrow\left(\sigma_{2},\:\sigma_{1}\right).
\end{align}
Therefore: 
\begin{align}
V_{c} & =r_{cs}^{L}\left(\sigma_{1},\:\sigma_{2}\right)\:I_{s,L}+r_{cs}^{R}\left(\sigma_{1},\:\sigma_{2}\right)\:I_{s,R},\\
-V_{c} & =r_{cs}^{L}\left(\sigma_{2},\:\sigma_{1}\right)\:I_{s,R}+r_{cs}^{R}\left(\sigma_{2},\:\sigma_{1}\right)\:I_{s,L},
\end{align}

which implies 
\begin{align}
r_{cs}^{L}\left(\sigma_{1},\:\sigma_{2}\right) & =-r_{cs}^{R}\left(\sigma_{2},\:\sigma_{1}\right).
\end{align}

Therefore for identical currents $I_{s,L}=I_{s,R}=I_{s}$
\begin{align}
V_{c} & =\left[r_{cs}^{L}\left(\sigma_{1},\:\sigma_{2}\right)-r_{cs}^{L}\left(\sigma_{2},\:\sigma_{1}\right)\right]\:I_{s}
\end{align}
which is antisymmetric under exchange of the magnetizations, $V_{c}\left(\sigma_{2},\:\sigma_{1}\right)=-V_{c}\left(\sigma_{1},\:\sigma_{2}\right).$
This implies a \textbf{parallel IMR} or vanishing response for parallel
magnetizations:
\begin{equation}
V_{c}\left(\sigma_{1},\:\sigma_{1}\right)=-V_{c}\left(\sigma_{1},\:\sigma_{1}\right)=0
\end{equation}
 and a bipolar effect for antiparallel magnetizations when flipping
both of them for identical currents: $V_{c}\left(\sigma_{1},\:-\sigma_{1}\right)=-V_{c}\left(-\sigma_{1},\:\sigma_{1}\right).$
Detailed calculations on an FNF trilayer exhibiting such a parallel
IMR will be given elsewhere. In the limit of infinite ferromagnets,
one finds in terms of a mean spin current $I_{s,0}^{m}=\left(I_{s,L}+I_{s,R}\right)/2$
and a spin current asymmetry $\delta I_{s}=\left(I_{s,L}-I_{s,R}\right)/2$:
\begin{equation}
MR=\frac{V_{c}^{AP}-V_{c}^{P}}{V_{c}^{P}}\sim\frac{I_{s,0}^{m}}{\delta I_{s}}
\end{equation}
(Length asymmetry drops out for infinite systems.)

\subsection{Anti-parallel IMR in CPP MTJ \& FNF trilayer.}

\begin{figure}[h]
\noindent \centering{}\includegraphics[width=1\columnwidth]{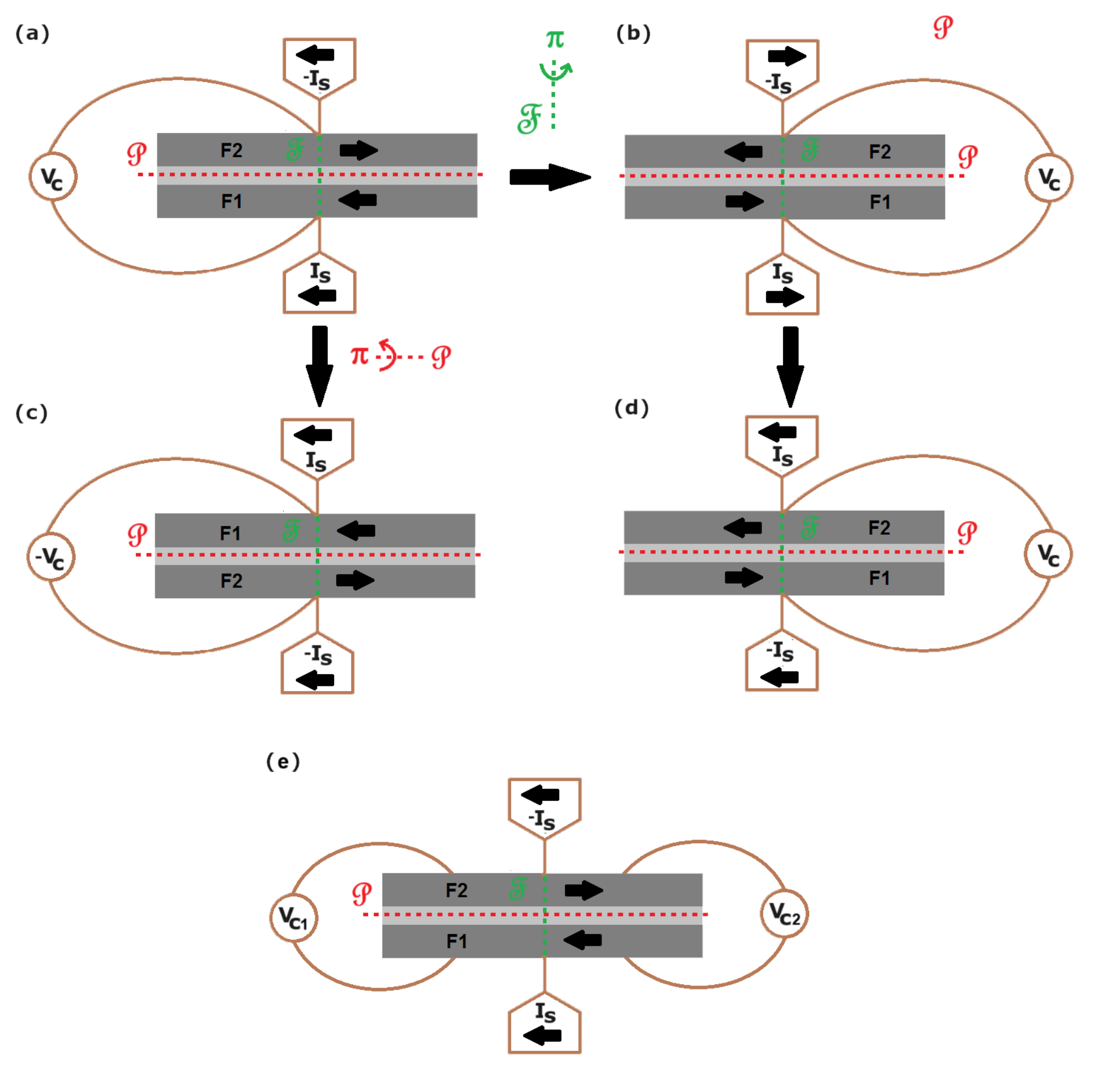}\caption{(Color online) (a) CPP symmetric spin valve connected to two spin
batteries injecting opposite spin currents $I_{s}$ into the ferromagnets.
The black arrows show the magnetization direction in each layer ($F1$or
$F2$) or the reference direction used to measure the spin current
(in the spin batteries). (b) Setup rotated around a symmetry axis
$\mathcal{D}$ assumed to exist. (b) Previous setup where the spin
currents take opposite values if the reference axis is flipped in
the spin batteries. (c) Setup rotated by an angle $\pi$ around a
symmetry axis $\mathcal{P}$ assumed to exist in the middle. (d) is
(b) with reference directions flipped in the spin batteries (which
changes the sign of spin currents). Comparing (c) and (d) shows they're
identical setups provided the ferromagnets $F1$ \& $F2$ are identical
(for symmetrical voltage probes), so that for anti-parallel magnetization,
there is no voltage drop across terminals $T_{1}$ and $T_{2}$. (e)
Instead of two voltage probes in the middle of the layers, it might
be easier to use 2 sets of voltage probes positioned symmetrically
with respect to axis $\mathcal{F}$. The previous discussion then
applies to $V_{c}=\left(V_{c1}+V_{c2}\right)/2$.\label{fig:cpp_2sb-1}}
\end{figure}

It is also possible to engineer an \textbf{anti-parallel IMR} (see
Fig.\ref{fig:imr-1} for the resistor analogy explaining it). Fig.\ref{fig:cpp_2sb-1}-a
shows such a setup. A CPP trilayer is connected symmetrically to two
spin batteries which inject opposite spin currents. If the voltage
probes are also symmetrical, the charge voltage must vanish as shown
in Fig.\ref{fig:cpp_2sb-1}. Indeed let us rotate the setup by an
angle $\pi$ around an axis $\mathcal{P}$ in the middle of the layers
(Fig.\ref{fig:cpp_2sb-1}-d): this exchanges the spin batteries and
the ferromagnets and the voltage drop gets reversed $V_{c}\longrightarrow-V_{c}$.
But one could also rotate the setup by an angle $\pi$ around an axis
$\mathcal{F}$ splitting the layers symmetrically (Fig.\ref{fig:cpp_2sb-1}-b):
all the magnetizations get reversed; the spin current reference axis
is also reversed but the charge voltage is obviously unchanged. But
a spin current $+I_{s}$ against a given magnetization reference axis
$\mathbf{M}$ is equivalent to a spin current $-I_{s}$ with a reversed
reference axis $-\mathbf{M}$. One therefore ends up with Fig.\ref{fig:cpp_2sb-1}-c
which is equivalent to Fig.\ref{fig:cpp_2sb-1}-d apart from an opposite
voltage. This implies that $V_{c}=0$. The obvious issue with such
a setup is that both the spin batteries and the voltage probes must
be located at the same positions. This is however easy to relax by
doubling the voltage probes and positioning them in a symmetric manner
with respect to symmetry plane $\mathcal{F}$ (see Fig.\ref{fig:cpp_2sb-1}-e);
one would then measure two voltages drops $V_{c1}$ and $V_{c2}$
which might not be offset-free but their average will be: $V_{c}=\left(V_{c1}+V_{c2}\right)/2$.

We illustrate such an AP-IMR with an MTJ (magnetic tunnel junction).
The junction voltage can be shown using Julliere model \citep{julliere_tunneling_1975}
and Slonczewski circuit theory \citep{slonczewski_conductance_1989}
to be:
\begin{equation}
V_{J}=-\frac{1}{g_{0}}\:\frac{P_{1}+P_{2}}{\left(1-P^{2}\right)^{2}}\:I_{s,J}
\end{equation}
where $P_{1/2}$ are the usual DOS polarizations, $P=\left|P_{1/2}\right|$,
$g_{0}=\Gamma\:n_{1}\:n_{2}$ (with $\Gamma$ the tunneling probability,
$n_{1/2}$ the total density of states) and $I_{s,J}$ the spin current
across the junction. This obviously vanishes for antiparallel magnetizations
which implies an \textbf{anti-parallel IMR}. This calculation neglects
however the spin accumulation effects in the ferromagnets. For very
thin ferromagnets, a more complete calculation incorporating the spin
accumulation in the ferromagnets, leads to 
\begin{equation}
TMR=\frac{V_{c}^{P}-V_{c}^{AP}}{V_{c}^{AP}}\sim\frac{P_{c}\,r_{c}}{P_{F}\,r_{F}\:\delta l}
\end{equation}
where $P_{c}\,r_{c}=\frac{1}{g_{0}}\:\frac{2\,P}{\left(1-P^{2}\right)^{2}}$;
$P_{F}$ is the conductivity polarization of the ferromagnets and
$r_{F}$ is their usual spin resistance; $\delta l=\left(d_{1}-d_{2}\right)/l_{F}$
is the difference in widths of the ferromagnets measured against the
spin relaxation length. Let us put some figures into it. For instance
the DOS polarization for Co is $P_{1}=P_{Co}\sim0.34$ at room temperature
in Moodera and coll.\citep{moodera_large_1995} so that $P_{c}=\frac{P_{1}+P_{2}}{1+P^{2}}=0.61$;
we also take $P_{F}\left(Co\right)\sim0.5$ (see for instance Ref.\citep{bass_cpp_2016})
and the ratio $r_{c}/r_{F}>10^{4}-10^{5}$. If we assume $l_{F}\left(Co\right)\sim40\:nm$
as in Table 3 of Ref.\citep{bass_spin-diffusion_2007} and that the
ferromagnet widths are equal up to a monolayer, $\delta l\sim0.1\,nm/40\,nm\sim2.5\:10^{-2}$
, then $TMR\sim10^{5}-10^{7}=10^{7}-10^{9}\%.$ The case of very thin
layers is the optimal one in terms of TMR ratio.

Let us consider layers which are not infinitely thin; for instance,
we assume equal widths $d_{1}=d_{2}=5\:nm$ up to a monolayer; this
means $l_{1/2}=d_{1/2}/l_{F}=0.125$. Using the same figures as above
numbers, a more realistic calculation (which will be detailed elsewhere)
shows that the spin current asymmetry can usually be neglected and
leads to $TMR\sim10^{5}\,\%$, which is still extremely large.

For an FNF trilayer in the same geometry, in the limit of thin layers,
one gets a similar MR ratio. The ratio is however much higher for
an MTJ because $r_{c}$ is considerably larger, by a factor $10^{3}-10^{5}$.
For all metallic spin valves, a spin battery based on ferromagnetic
resonance would be able to achieve signals in the $1-100\:\mu V$
range. By using an MTJ, since the resistance is multiplied by a factor
$10^{3}-10^{5}$, this implies that $mV$ signals are very achievable
which would make such a spin battery driven MTJ quite useable.

What would be the impact of MgO barriers ? This would increase the
value of $P_{c}$ and the TMR ratio. But it is not as critical as
having very small widths for the ferromagnets as well as a very small
width asymmetry. For instance, let us consider figures extracted from
Yuasa and coll.\citep{yuasa_giant_2004}. $RA\sim10-10^{7}\Omega\mu m^{2}$
and the TRM ratio reaches $180\:\%$ so $P_{1}=0.69$ or $P_{c}=0.93$.
If we plug this in the previous TMR ratio calculation, this would
increase it by a factor 1.5 which is nice but not as crucial as having
a very small $\delta l$.

\section{Conclusion. }

We have introduced several spin valve setups which due to symmetry
can exhibit both the bipolar effect of Johnson spin transistor and
an IMR. An obvious application of IMR and BE would be more sensitive
magnetic sensors so an experimental confirmation of these predictions
would be very useful. An essential component for the occurence of
both BE and IMR is to replace charge batteries by spin batteries since
this allows the occurence in response functions of odd terms $\propto sgn\left(M_{z}\right)$.
Since real systems are never perfectly symmetric, we have also quantified
the impact of length \& spin current asymmetries. The MR ratio can
still reach high ranges $10-10^{3}$. The simplest system to observe
the BE is the twin system with shared spin battery; for observation
of IMR, it is better to test on a twin system with independent spin
batteries. The CPP MTJ with two spin batteries is quite promising
in comparison with the CPP FNF trilayer due to the much larger HIGH
signal which should make the $mV$ range quite reachable. Extensions
of this work to non-collinear systems with symmetries, to CIP setups,
to parallel IMR in CPP systems and inclusion of non-ideal spin batteries
are discussed elsewhere. 

\bibliographystyle{apsrev4-2}
\bibliography{jpsjspin}

\end{document}